\documentclass{frontiersSCNS} 

\usepackage{url,lineno,natbib}

\copyrightyear{2014}
\pubyear{2014}

\def\journal{Solar and Stellar Physics}
\def\DOI{}
\def\articleType{Research Article}
\def\keyFont{\fontsize{8}{11}\helveticabold }
\def\firstAuthorLast{McIntosh {et~al.}} 
\def\Authors{Scott W. McIntosh\,$^{1,*}$, Robert J. Leamon\,$^{2}$}
\def\Address{$^{1}$High Altitude Observatory, National Center for Atmospheric Research, Boulder, CO, USA\\
$^{2}$Department of Physics, Montana State University, Bozeman, MT, USA}
\def\corrAuthor{Scott W. McIntosh}
\def\corrAddress{High Altitude Observatory, National Center for Atmospheric Research, P.O. Box 3000, Boulder, CO, 8030, USA}
\def\corrEmail{mscott@ucar.edu}

\newcommand{\pref}{\protect\ref}
\newcommand{\soho}{{\em SOHO{}}}
\newcommand{\sdo}{{\em SDO{}}}
\newcommand{\degree}{${^\circ{}}$}

\begin{document}
\onecolumn
\firstpage{1}

\title[Grand Minima]{Deciphering Solar Magnetic Activity: On Grand Minima in Solar Activity}
\author[\firstAuthorLast ]{\Authors}
\address{}
\correspondance{}
\extraAuth{}
\topic{Solar and Stellar Physics}

\maketitle

\begin{abstract}
\section{}
The Sun provides the energy necessary to sustain our existence. While the Sun provides for us, it is also capable of taking away. The weather and climatic scales of solar evolution and the Sun-Earth connection are not well understood. There has been tremendous progress in the century since the discovery of solar magnetism \--- magnetism that ultimately drives the electromagnetic, particulate and eruptive forcing of our planetary system. There is contemporary evidence of a decrease in solar magnetism, perhaps even indicators of a significant downward trend, over recent decades. Are we entering a minimum in solar activity that is deeper and longer than a typical solar minimum, a "grand minimum"? How could we tell if we are? What is a grand minimum and how does the Sun recover? These are very pertinent questions for modern civilization. In this paper we present a hypothetical demonstration of entry and exit from grand minimum conditions based on a recent analysis of solar features over the past 20 years and their possible connection to the origins of the 11(\--ish) year solar activity cycle.
\tiny
\keyFont{ \section{Keywords:}dynamo \-- convection \-- magnetic fields \-- sunspots \-- Sun: evolution \-- Sun: activity \-- Sun: helioseismology \-- Sun: interior \-- Sun: fundamental parameters \-- stars: activity \-- stars: evolution}
\end{abstract}

\section{Introduction}
We live in the outer atmosphere of our star. As a result, the electromagnetic radiation from our star is an essential ingredient in our day-to-day existence and, as our civilization becomes increasing dependent on technology, we become increasingly susceptible to our star's occasional tantrums. The magnetism of our star shapes the interaction with our (magnetically shielded) planet on the largest (magnetic) scales while generating the radiative and particulate outputs that bathe our atmosphere on the smallest \citep[e.g.,][]{2013ApJ...765..146M, Whomps}. Over the past 30 years our awareness of this systemic interaction has grown, and we have begun to notice that the radiative and particulate forcing of our atmosphere is slowly decaying \--- likely driven by a downturn in the Sun's global magnetism. It is unclear what effect this gross downturn in solar activity will have on our planet if it continues to trend downward. Are we entering a new grand minimum? Our understanding of such ``grand minima'' in solar activity is scant. How they arise, and how the Sun recovers from them, is a difficult proposition to undertake just due to their infrequency, but a body of literature exits pursuing a variety of theoretical models \citep[e.g.,][]{2006AdSpR..38..856G, 2008SoPh..250..221M, 2008AN....329..351B, 2010ApJ...724.1021K}. However, we can also attribute our poor understanding of the grand minimum state to the fact that our understanding of the process(es) that generate the Sun's magnetic field and make it cycle are relatively poor, despite very active research \citep[e.g.,][]{2010LRSP....7....3C}. In the next few decades it will be critical to assess the forcing of the planet and solar system, as we may need to cope with potentially catastrophic climatic evolution and also begin to send human explorers outward from home to other planets and celestial bodies.  

\begin{figure}
\begin{center}
\includegraphics[width=\hsize]{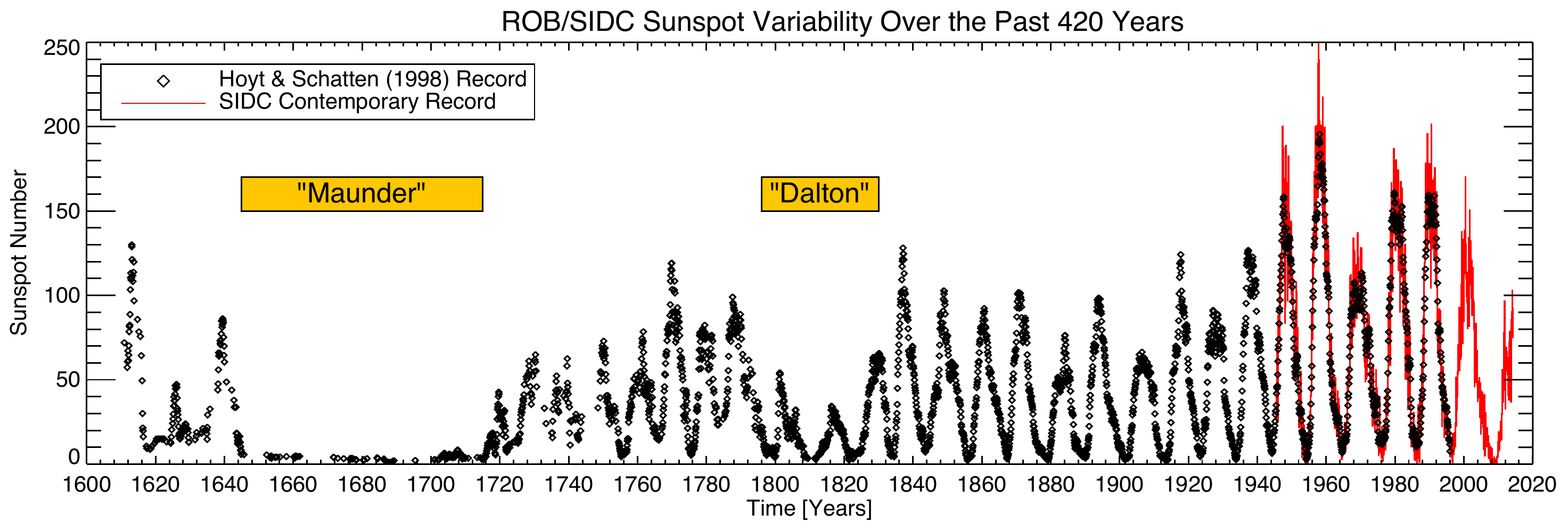}
\end{center}
\textbf{\refstepcounter{figure}\label{f1} Figure \arabic{figure}.}{ Variations in total sunspot number over the past 400+ years. In this record we see the prevalent $\sim$11-year quasi-periodicity of the solar cycle in addition to possible, longer period, envelopes of activity. Two of the most prevalent downturns in sunspot activity over this epoch were dubbed the ``Maunder'' and ``Dalton'' minima, so-called grand minima. Source: WDC-SILSO, Royal Observatory of Belgium, Brussels}
\end{figure}

Beyond the canonical 11(-ish) year solar activity cycle, the Sun's variability appears to undergo long-term modulation with characteristic periods of 80, 150, 400 years, and possibly even longer \citep[e.g., ][]{lrsp-2013-1}. Understanding those variations offers us an opportunity to understand potential connections between the Sun and terrestrial climatic changes on similar timescales. By far the most dramatic of these longer-term cycling changes is the notion of a grand minimum \--- like the ``Maunder Minimum'' of the seventeenth century where sunspots effectively disappeared for 75 years \citep[see, e.g., ][and Fig.~\pref{f1}]{1976Sci...192.1189E}. Interestingly, while sunspots all but disappeared from view, other markers of solar activity continued to cycle with a period of nearly 11 years. For example, radio-isotope records \citep[][]{1998SoPh..181..237B} show slightly longer periods (than 11 years) during the deep minimum and a couple of cycles of shorter periods on exit before the sunspots return almost perfectly in phase at the end of the grand minimum.

Some recent insight into the origins of the $\sim$11-year quasi-periodicity of the solar activity (or sunspot) cycle was presented in \cite{2014ApJ...792...12M} and provides the motivation for the discussion to follow.

\section{Results}
%

In this section we discuss the concept of magnetic activity band interaction in the context of solar cycle 23 as originally discussed by \cite{2014ApJ...792...12M} and use the phenomenological description of solar cycle modulation described therein to present a hypothetical progression of our Sun into (and out of) grand (sunspot) minima. We see that this hypothetical progression inherently preserves cyclic variability of global magnetism while significantly reducing the occurrence of sunspots for an extended period of time before returning to a normal 11-ish year sunspot cycle. Finally, we compare the observed variation of signposts over the past 70 years with the hypothetical model to illustrate where we {\it may} be in the progression to grand minimum should the apparent downturn in solar magnetism continue.

\subsection{Activity Band Interaction and The Sunspot Cycle}
\cite{2014ApJ...792...12M} presented a possible explanation of the 11(-ish) year solar cycle in the context of the spatio-temporal evolution of coronal brightpoints and what appear to be the markers of the giant convective scale \citep[][]{2014ApJ...784L..32M}. Using observations from \soho/EIT, \soho/MDI, \sdo/AIA and \sdo/HMI that span from 1996 to 2014, they deduced that the landmarks and phases of the 11-ish year sunspot cycle: the ascending phase, solar maximum, the declining phase, and solar minimum arise as a result of the latitudinal-temporal interaction of (toroidal) magnetic flux systems (or what we will refer to hereafter as ``activity bands'') that belong to the 22-year magnetic activity cycle. The bands of magnetic activity in each hemisphere of the same polarity appear with tremendous regularity, which \cite{2014ApJ...792...12M} coined as a high latitude ``clock'' for each hemisphere. In short, the appearance of the sunspot cycle could be explained in terms of the destructive ``interference'' of these oppositely signed activity bands as they migrate equatorward (Fig.~\pref{f2}), with the speed of the migration impacting the strength of the sunspot cycle (fast \-- high; slow \-- low). 

The main result of \cite{2014ApJ...792...12M} is illustrated in panel B of Fig.~\pref{f2}. The temporal and latitudinal interaction of the oppositely signed activity bands in each hemisphere, and across the equator, modulate the occurrence of sunspots (on the low-latitude pair). \cite{2014ApJ...792...12M} noticed that the activity bands of same sign appear at high latitudes ($\sim$55\degree) every 22 years and migrate equatorward, taking approximately 19 years to reach the equator. The low-latitude pair then abruptly ``terminate'', for example the cycle 23 sunspots did not appear to grow in abundance or size until the cycle 22 bands had terminated (in mid-1997). Similarly, the polarity mirror-image of this progression occurred in early 2011 for cycle 24 sunspots \--- following the termination of the cycle 23 bands. This equatorial termination, or cancellation, signals the end of one sunspot cycle and leaves only the higher-latitude band in each hemisphere. Sunspots rapidly appear and grow on that band for several years until the next (oppositely-signed) band appears at high latitude---defining the maximum activity level of that new cycle and triggering a downturn in sunspot production. The perpetual interaction of these temporally offset 22-year activity bands appears to drive the quasi-11-year cycle of sunspots. This progression, by definition, forms the envelope of the Sun's magnetically driven activity where the degree of overlap in the bands appears to inversely govern our star's sunspot production (more overlap, less spots and vice versa). This observational evidence presented by \cite{2014ApJ...792...12M} points to the rotational energy at the bottom of our Star's convective interior as being a/the major driver of its long-term (magnetic) evolution.

\subsection{Grand Minimum Hypothesis}
We now present an extension of the activity band overlap and interaction concept discussed by \cite{2014ApJ...792...12M}. While a detailed physical description of this coupled system is not yet in place, we can use insight developed above to explore whether our paradigm can place grand minima of solar activity in context. Can the Sun go through epochs of (very) low sunspot production discussed in \cite{1976Sci...192.1189E} and still produce the signature of global modulation and magnetic cycling evidenced in the radionuclide measurements \citep[][]{1998SoPh..181..237B} without appearing to miss a beat when the sunspot cycle returns? It would appear that the Sun can do this ``grand minimum trick'' without magically switching the dynamo process off! 

From the historical analysis presented in \cite{2014ApJ...792...12M} we see that weaker sunspot cycles are generally longer. That is, in the band-overlap picture, the magnetic activity band interaction occurs over a longer period of time \--- inhibiting sunspot formation for a longer spell. Further, they noticed that one weak cycle tends to beget another, where systematic weakening of the subsurface magnetic field slows the global circulation pattern and increases the overlap time of subsequent cycles. The extended overlapping, interaction, and mutual cancellation of sub-surface flux systematically reduces sunspot production. Could this process continue until there is complete mutual cancelation of the activity bands that results in no sunspot formation for an extended period of time? And, could this sunspot hiatus come to a natural end that would result in the sunspot cycle returning to ``normal''?

In the band-interaction paradigm the critical issue appears to be related to the extension of the equatorial band termination time. The contemporary state of the sunspot cycle sees a time of a couple of years between termination and the maximum of the next cycle. We suppose that decreasing the time that the single hemispheric band (in each hemisphere) has to grow sunspots post-termination will result in significantly less sunspots being formed in each hemisphere and in total. This appears to happen when the progression from high to low latitudes is slow and there is extended period of overlap between the four bands (two per hemisphere). Now, such behavior would persist until the 22-year high latitude ``clock" (critical to the progression of the band interaction and the formation of the sunspot cycle) naturally introduces a third activity band in each hemisphere. The asymmetry that this third band introduces would immediately start to increase the amount mutual band interaction. It is highly likely that this complex three/six-band interaction would produce very complex sunspot patterns that may have been observed at the end of the Maunder Minimum \cite[][]{2009SoPh..255..143A}. The interplay of the three/six band system, while it persisted, would likely result in shorter (8-9 year) cycles being observed \citep[e.g.,][]{1998SoPh..181..237B} and likely a speed-up of the latitudinal progression. The mutual cancelation in the three/six band system would almost certainly revert to a two/four band configuration in relatively short order.

\begin{figure}
\begin{center}
\includegraphics[width=\hsize]{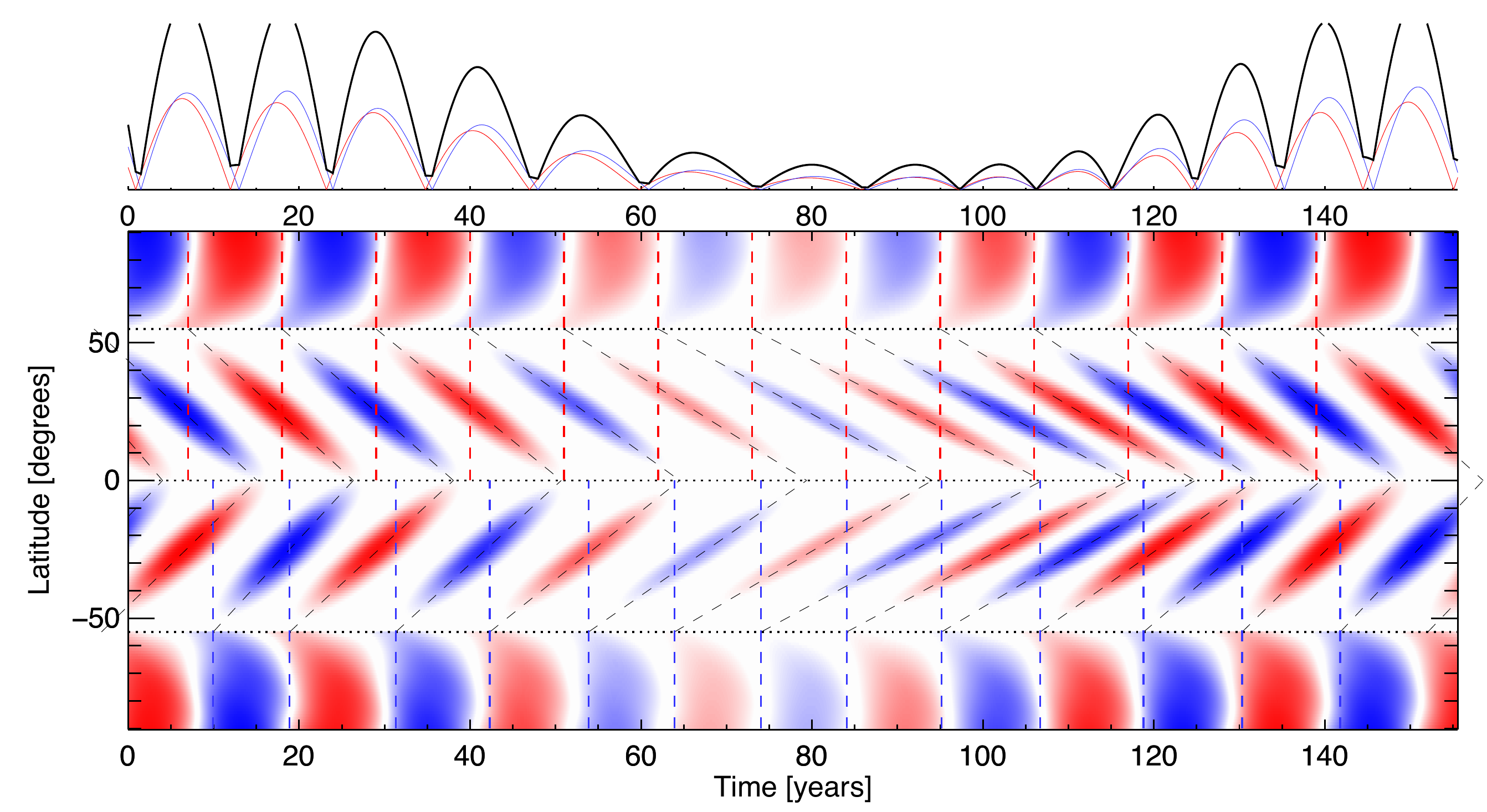}
\end{center}
\textbf{\refstepcounter{figure}\label{f3} Figure \arabic{figure}.}{ Representing the progression of solar cycle variation into, and out of, a grand minimum state based on the simple metrics developed above. The dashed vertical lines mark the onset of the equator-ward branch in the northern (red) and southern (blue) hemisphere.}
\end{figure}

Figure~\pref{f3} shows a representation of how the solar cycle can transition into a grand minimum state and out again without abrupt changes to the underlying circulation or switching off the dynamo. For this illustration the northern hemisphere has a fixed 22-year period, while the southern hemisphere has a 22-year period that is randomly perturbed in a $\pm$2-year range to reduce the north-south symmetry\footnote{Note that we have included a fixed two year offset between the hemispheres for illustrative purposes. The southern 22-year activity bands following two years behind those of the north in the start of their equatorial migration. It is unlikely that a phase offset like this, or that presently ongoing, is a necessary condition for the onset of grand minimum conditions.}. The termination points of the equatorial branches are then stretched out from 20 years using a Gaussian perturbation 30 years wide that is centered on 75 years before they return to their normal spacing. The top panel of the figure shows the integrated latitudinal signal in each hemisphere and the total (black line) as a proxy of the number of sunspots. 

At the start of the sequence, before we start to increase the termination points, there are 11-year cycles in the system. When the termination points are extended we see a progression from 11\-- to 13\-- (and one 14) year cycles where the latter occur when the termination points are maximally extended and before the high latitude variation introduces a third activity band (about year 85). After this time the activity bands in each hemisphere appear to ``pile up'' as the system begins to come out of the deep minimum. This pile-up, and rapid cancellation, produces a couple of short cycles (periods around 9 and 10 years) before the 11-year variation is re-established near the end of the sequence. 

This progression of cycle lengths is consistent with that presented in Table 1 of \cite{1998SoPh..181..237B} for the Maunder Minimum and is driven be the slow-down of the high to low latitude band progression and its interaction with the high latitude clocks in each hemisphere.

\subsection{Where Are We Now?}
\begin{figure}
\begin{center}
\includegraphics[width=\hsize]{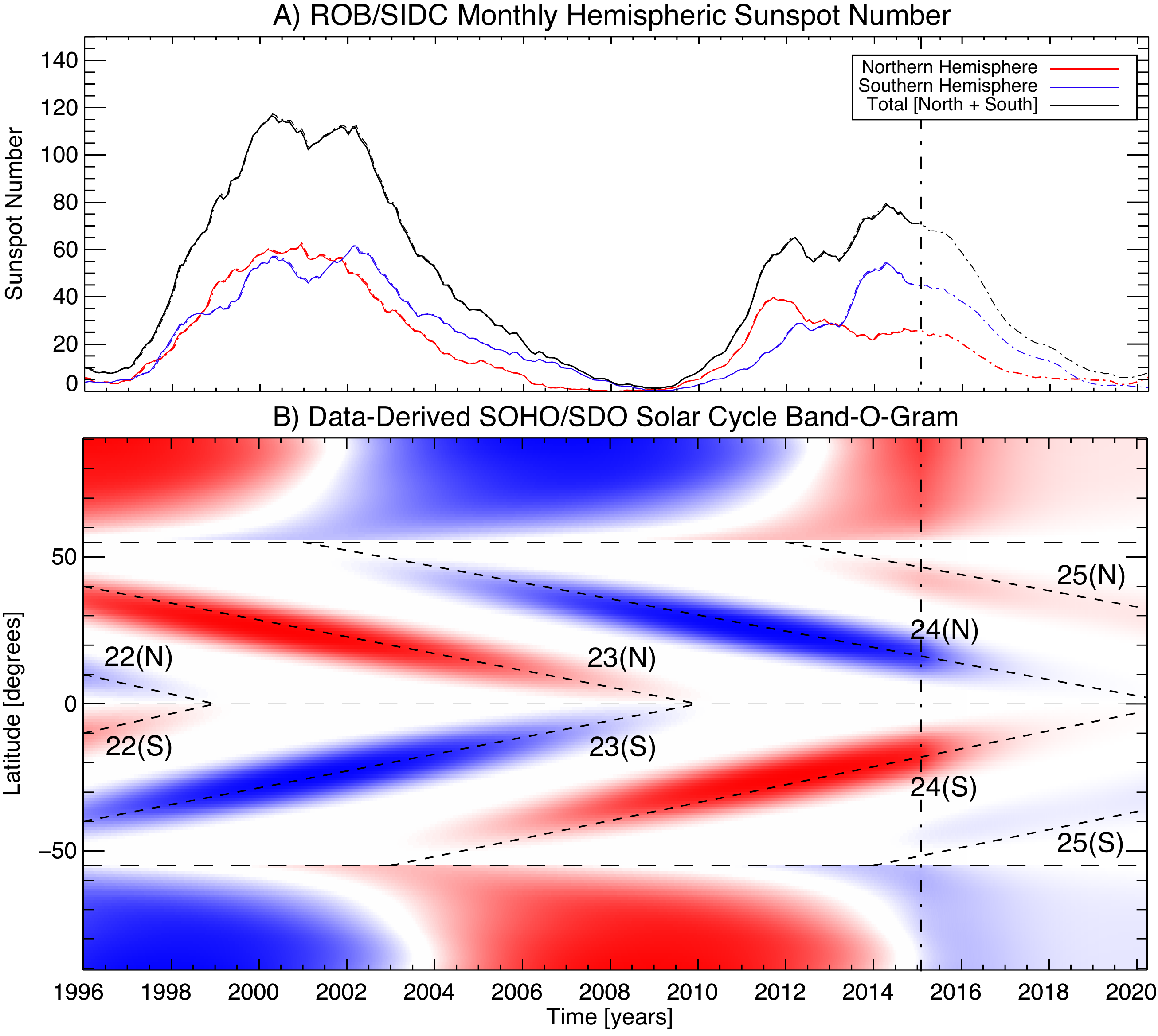}
\end{center}
\textbf{\refstepcounter{figure}\label{f2} Figure \arabic{figure}.}{ Panel A shows the total and hemispheric sunspot numbers for cycles 23 and 24 as black, red and blue solid lines, respectively. Panel B shows the data-determined schematic of the migrating activity bands that belong to the 22-year magnetic activity cycle from \citet{2014ApJ...792...12M}. The interplay of these bands gave rise to sunspot cycles 22, 23, and 24 as indicated. The reader will notice that, in addition to the cycle 25 bands appearance between 2012 and 2014, we have attempted to indicate possible future behavior of the system as outlined in \citet{2014ApJ...792...12M} and is shown with an opacity mask. In each case the color of the band reflects its polarity (red--positive; blue--negative). The dashed horizontal lines drawn indicate the equator and lines of the 55th parallel in each hemisphere while the vertical dot-dashed lines indicate the current time. Source: WDC-SILSO, Royal Observatory of Belgium, Brussels}
\end{figure}

On the basis of the slow downturn in sunspot production over the past three decades, is the contemporary Sun on the cusp of a new grand activity minimum? Before we address this question we show the variation in the total (and hemispheric) sunspot numbers over the past 70 years in Fig.~\pref{f2}. This plot illustrates some of the points made by \cite{2013ApJ...765..146M} and adds further context to Figs.~\pref{f1} and~\pref{f2} where we see the exchange of dominance between the northern and southern hemispheres in addition to the general downturn of both the total and hemispheric sunspot numbers over the past 25 years or so. 

\begin{figure}
\begin{center}
\includegraphics[width=\hsize]{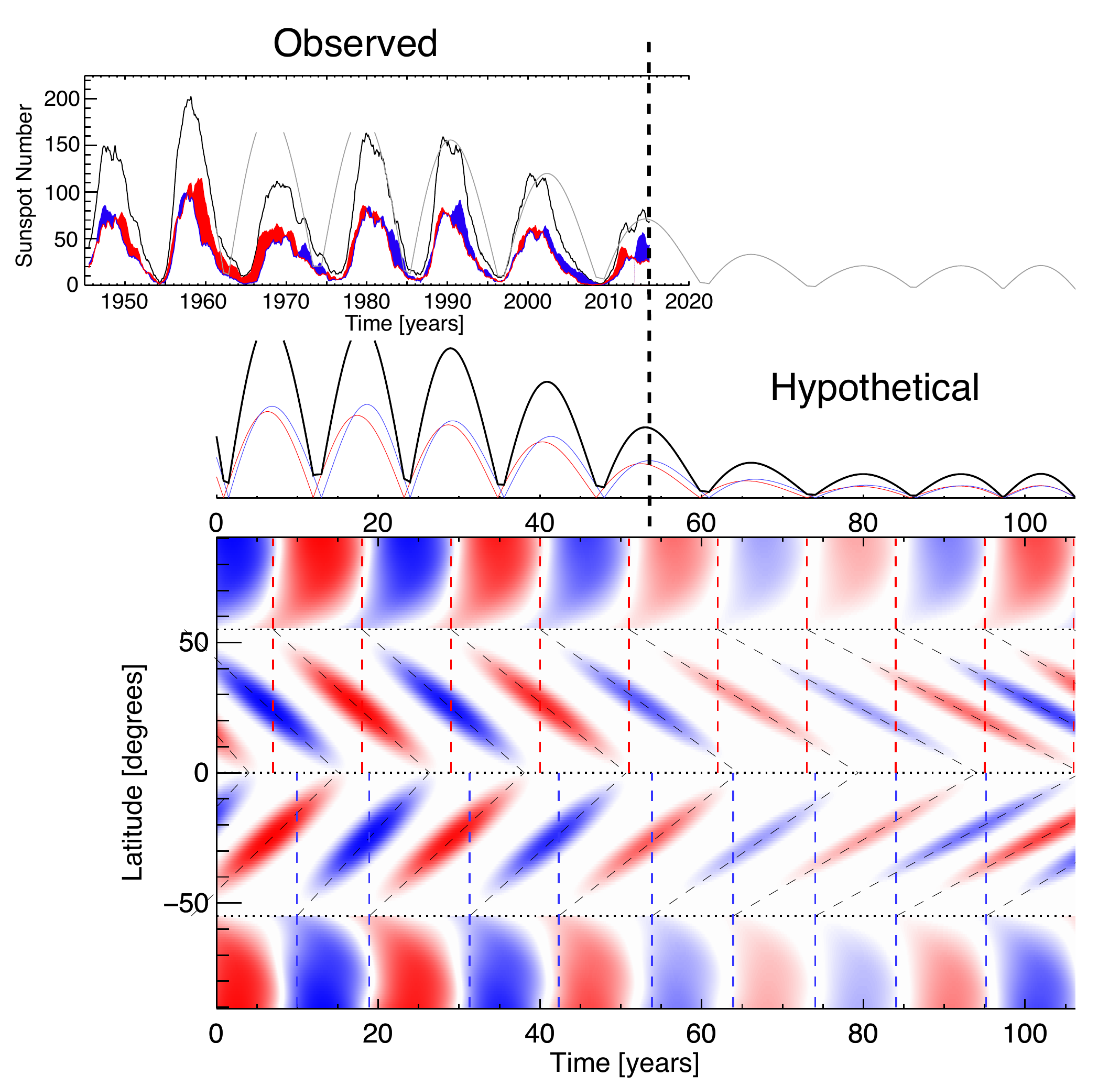}
\end{center}
\textbf{\refstepcounter{figure}\label{f5} Figure \arabic{figure}.}{ Combining observation (Fig.~\pref{f2}) and our schematic progression into grand minima (Fig.~\pref{f3}) to visually compare the progression of sunspot number variation and qualitatively assess where we might be in a general decline of activity if the downward trend demonstrated over the past couple of decades were to continue based on our earlier simple assumptions. In the top panel, for visual comparison we show the cartoon progression of the total sunspot number as a thick dashed gray line.}
\end{figure}

Fig.~\pref{f5} blends the observations of Fig.~\pref{f2} (top) with the conceptual progression into grand minimum motivated by the band-overlap and interaction paradigm of Fig.~\pref{f3} (bottom). We simply use the hypothetical envelopes of the total and hemispheric sunspot to assess whether the current decline carries {\em any} information about where we may be in the process of grand minimum entry {\em if} the Sun continues its downward trend. While this is not a rigorous scientific comparison of observation and model, we see that there is a strong visible comparison between the observations and conceptual model from years 30 to 55, where we appear to be presently (vertical thick dashed line). Naturally, the Sun often throws us curveballs, which illustrate our lack of understanding or profound ignorance. However, based on the current observations and the model motivated by band-overlap from the contemporary data, we are around 35 years from the maximal cancellation and reduction in sunspot production. It seems like a new extended minimum in sunspot production {\em may} be upon us, we should monitor closely for additional slow-down of the band migration that would result in a lengthening of the overlap period and mutual cancellation that would lead into a grand minimum state.

\section{Discussion}
In the figures and material presented above we have discussed the potential of the Sun to enter and exit a grand minimum of solar activity based on the concept of the overlap and interaction of magnetic activity bands that belong to our star's 22-year magnetic polarity cycle. In short, the degree of intra- and extra-hemispheric overlap between bands of opposite polarity appears to moderate the production of the sunspots belonging to the 11-ish year sunspot cycle --- the greater the overlap, the greater the cancelation, the smaller the resulting sunspot number. Therefore, we see that the key factor that appears to control the amplitude of the sunspot number is the length of time that it takes for these magnetic bands to migrate from high to low latitudes \--- slow migration increases overlap time, fast migration reduces it \citep[][]{2014ApJ...792...12M}. With a very simple phenomenological model we have demonstrated that a progression into, and out of, grand minima can be accomplished intuitively via a systematic increase in overlap time that eventually results in global magnetic asymmetry that re-establishes the progression of the 11-ish year cycle. We see that during this increase in band overlap there would be very few sunspots for a prolonged period of time, while the global magnetic field will continue to cycle, consistent with inferences from radionuclide analysis \citep[e.g.,][]{1998SoPh..181..237B}. Finally, by comparing this conceptual demonstration with contemporary observations, there is a hint that the decline in solar magnetism, and other activity proxies observed over the past 25 years \citep[e.g.,][]{2013ApJ...765..146M}, could be indicative of a decline into a modern minimum in solar activity. Using markers of this global\--scale magnetic evolution \citep[][]{2014ApJ...784L..32M} we can continue to trace this evolution while a physically consistent model of the magnetic evolution can be constructed.

\section*{Disclosure/Conflict-of-Interest Statement}
The authors declare that the research was conducted in the absence of any commercial or financial relationships that could be construed as a potential conflict of interest.

\section*{Author Contributions}
The authors played an equal role in the original data analysis, writing, presentation and editing of the material discussed.

\section*{Acknowledgement}
The data used in this paper are openly available from the {\em SOHO}, {\em SDO}, and the Virtual Solar Observatory (VSO; \url{http://virtualsolar.org}) data archives. {\em SOHO} is a project of international collaboration between ESA and NASA. Sunspot data were provided by David Hathaway and the World Data Center SILSO, Royal Observatory of Belgium, Brussels.
\paragraph{Funding\textcolon} NCAR is sponsored by the National Science Foundation. We acknowledge support from NASA contracts NNX08BA99G, NNX11AN98G, NNM12AB40P, NNG09FA40C ({\em IRIS}), and NNM07AA01C ({\em Hinode}).

\bibliographystyle{frontiersinSCNS&ENG} 
\bibliography{mcintosh}

\section*{Figures}

\begin{figure}
\begin{center}
\includegraphics[width=\hsize]{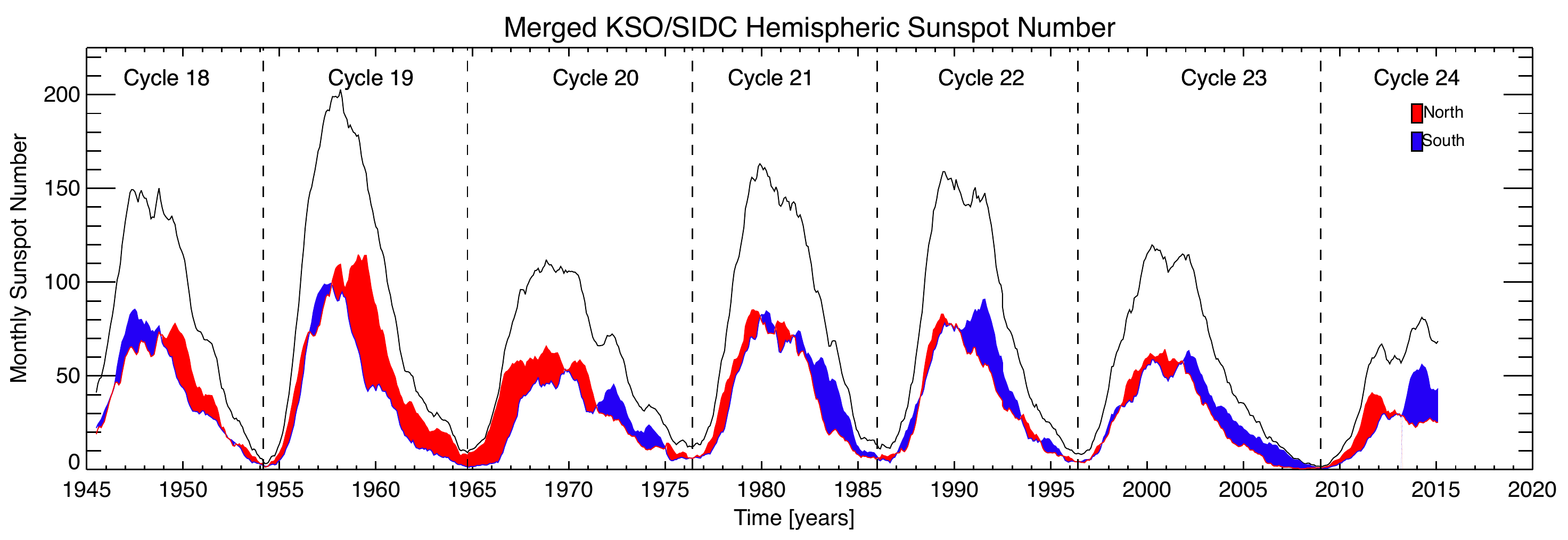}
\end{center}
\textbf{\refstepcounter{figure}\label{f4} Figure \arabic{figure}.}{ The international sunspot number by hemisphere over the past seventy years. Constructed from data provided by the Kanzelh\"{o}he Solar Observatory (KSO) and the ROB/SIDC from 1945 to the present. The northern (red) and southern (blue) hemispheric sunspot numbers are shown against the total of the two (black). The shading of the difference between the northern and southern sunspot numbers indicates an excess in sunspot number between them. The dashed vertical lines are drawn at approximate sunspot minima to delineate the cycles over the timeframe. Source: WDC-SILSO, Royal Observatory of Belgium, Brussels}
\end{figure}

\end{document}